\documentclass[]{article}
\usepackage{graphicx}                    
\usepackage[square]{natbib}
\begin{document}
\title{Type II and IV radio bursts in the active period October-November 2003}

\author{V. Petoussis,P. Tsitsipis,A. Kontogeorgos\\
\emph{Technological Education Institute of Lamia, Lamia, Greece}\\
X. Moussas,P. Preka--Papadema, , C. Caroubalos, A. Hillaris\\
\emph{University of Athens, 15783 Athens, Greece}\\
C. E. Alissandrakis\\ \emph{University of Ioannina, 45110 Ioannina, Greece}
J.-L. Bougeret, G. Dumas\\ \emph{Observatoire de Paris, CNRS UA 264, 92195 Meudon Cedex, France}}
\maketitle
\begin{abstract}
In this report we present the Type II and IV radio bursts observed and analyzed by the radio spectrograph 
ARTEMIS IV\footnote{Appareil de Routine pour le Traitement et l' Enregistrement Magnetique de l' Information Spectral}, 
in the 650-20MHz frequency range, during the active period October-November 2003. These bursts exhibit very rich fine 
structures such fibers, pulsations and zebra patterns which is associated with certain characteristics of the associated 
solar flares and CMEs.
\end{abstract}

\section{Introduction}
The active period from 20 October to 5 November 2003 has attracted considerable attention 
(\cite{Chertok}, \cite{Kiener}, \cite{Mitsakou}, \cite{Tan}, \cite{Gopalswamy}, \cite{Vrsnak}, \cite{Wang}) because
of its abundance in of powerful flares and large coronal mass ejections (CMEs) as well as by strong space weather 
disturbances. This activity was associated with the appearance of a global complex consisting of three large, 
remote but connected active regions (AR): AR 484 (Carrington coordinates, N04; L = 354), AR 486 (S15; L = 283) and
AR 488 (N08; L = 291) (cf. \cite{Chertok}); while in the visible hemisphere of the sun, 
this complex, produced 8 X--class and 12 M--class flares.

The Artemis IV solar radio-spectrograph operating at Thermopylae since 1996 (\cite{Caroubalos01a})
consists of a 7--m parabolic antenna covering the metric range, to which a dipole antenna was added recently in
order to cover the decametric range (\cite{Caroubalos06}, \cite{Kontogeorgos}). 
Two receivers operate in parallel, a sweep frequency analyser (ASG) covering
the 650--20 MHz range in 630 data channels with a cadence of 10 samples/sec and a high sensitivity multichannel
acousto--optical analyser (SAO), which covers the 270--450 MHz range in 128 channels with a high time
resolution of 100 samples/sec.
\begin{figure}[h!]
\begin{minipage}[t]{6cm}
\begin{center}
\includegraphics[width=6cm]{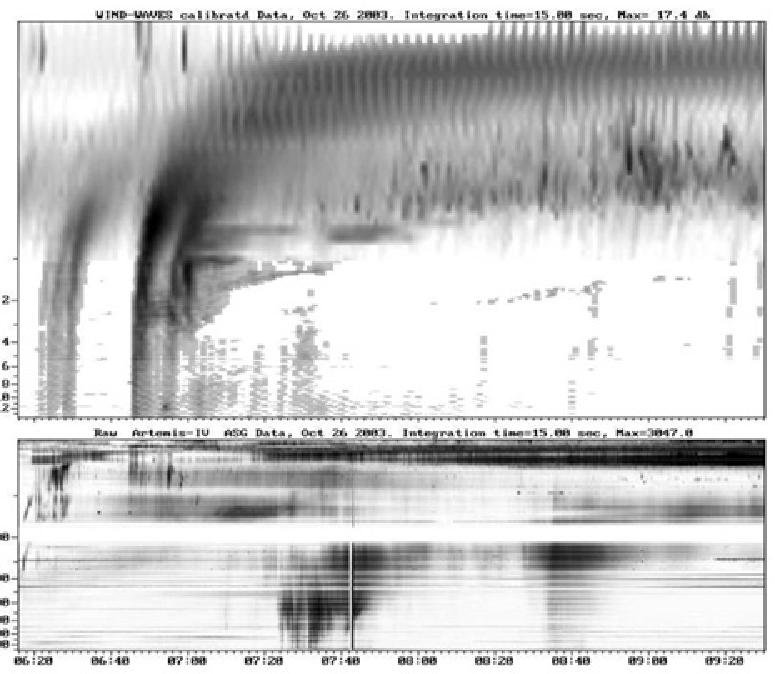}
\end{center}
\end{minipage}
\hfill
\begin{minipage}[t]{6cm}
\begin{center}
\includegraphics[width=6cm]{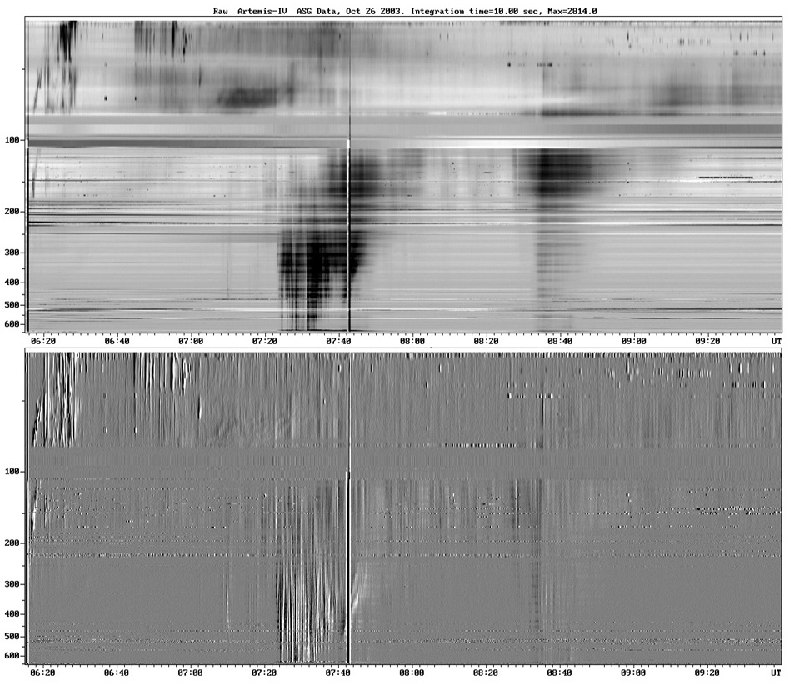}
\end{center}
\end{minipage}
  \caption{LEFT:Combined ARTEMIS-IV and WIND--WAVES spectra of the October 26, 2003 event. 
RIGHT: ARTEMIS--IV pectra in the frequency range 20--650 MHz; Intensity (top)and its time derivative (bottom), 
enhancing fine temporal details}
 \label{03OCT26_01}
 \end{figure}
\begin{figure}[h!]
\begin{minipage}[t]{6cm}
\begin{center}
\includegraphics[width=6cm]{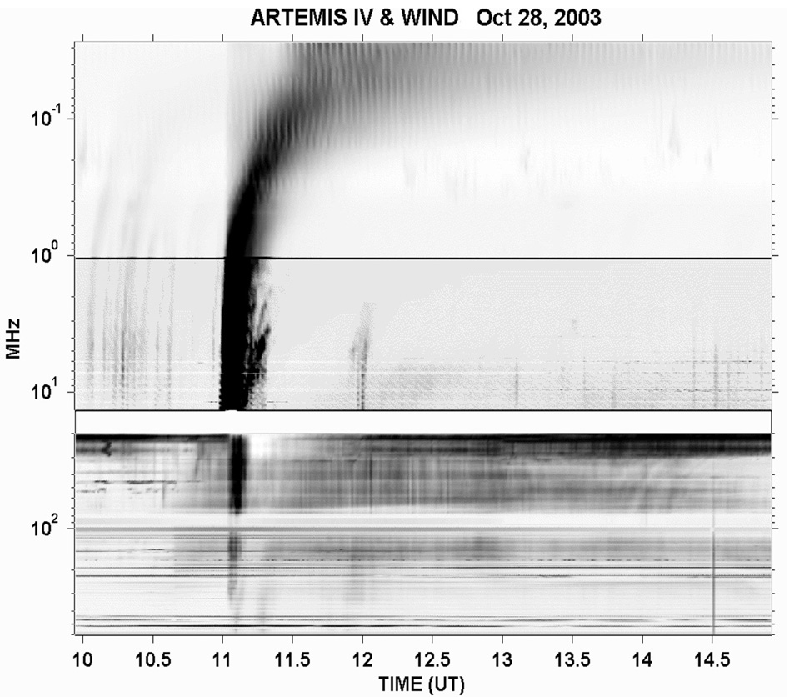}
\end{center}
\end{minipage}
\hfill
\begin{minipage}[t]{6cm}
\begin{center}
\includegraphics[width=6cm]{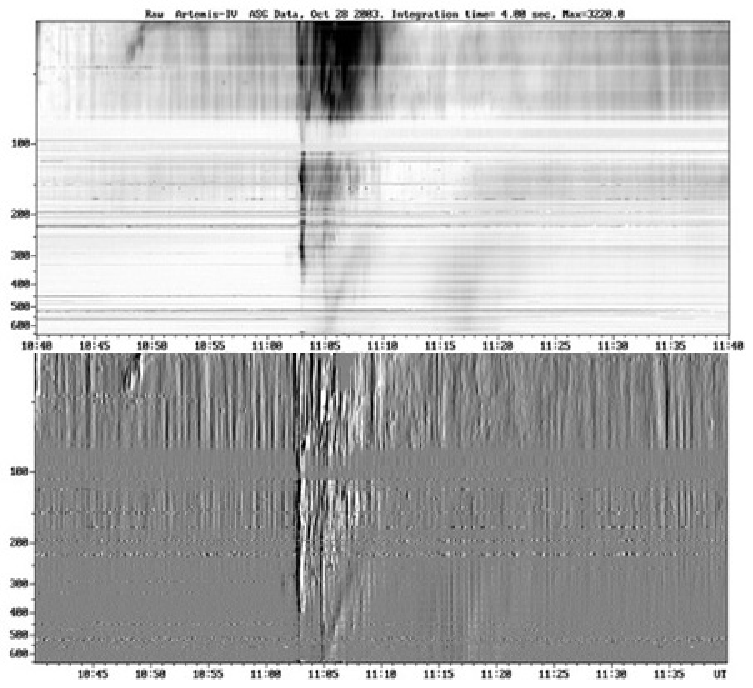}
\end{center}
\end{minipage}
  \caption{LEFT: Combined ARTEMIS-IV and WIND--WAVES spectra of the October 28, 2003 event.
RIGHT: ARTEMIS--IV spectra in the frequency range 20--650 MHz; Intensity (top)and its time derivative (bottom), 
enhancing fine temporal details }
 \label{03OCT28_01}
 \end{figure}
\begin{figure}[h!]
\begin{minipage}[t]{6cm}
\begin{center}
\includegraphics[width=6cm]{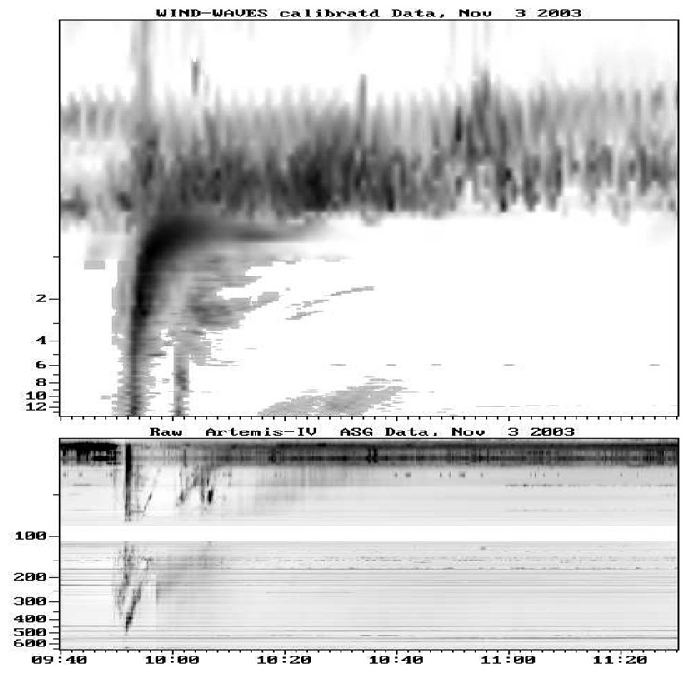}
\end{center}
\end{minipage}
\hfill
\begin{minipage}[t]{6cm}
\begin{center}
\includegraphics[width=6cm]{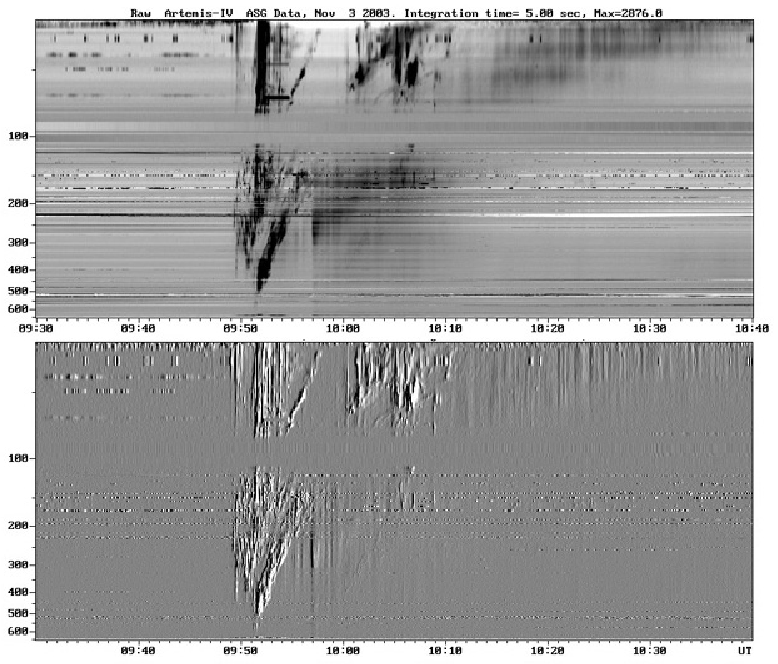}
\end{center}
\end{minipage}
\centerline{\includegraphics[width=\textwidth,height=3cm]{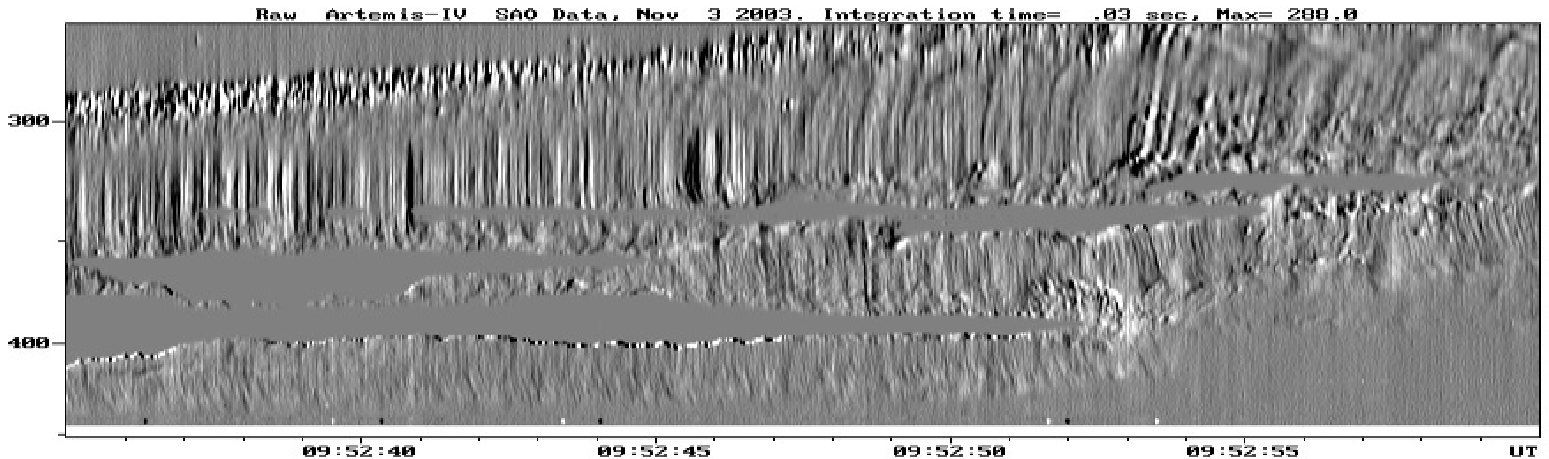}}
  \caption{UPPER LEFT: Combined ARTEMIS-IV and WIND--WAVES spectra of the November 3, 2003 event.
UPPER RIGHT: ARTEMIS--IV ASG Spectra in the frequency range 20--650 MHz; 
Intensity (top)and its time derivative (bottom), enhancing fine temporal details.
BOTTOM PANEL: Differential SAO Spectrum with high (0.03 sec) resolution of a part Type II lane, showing
U--shaped \emph{herringbone} structures.}
 \label{03NOV03_01}
 \end{figure}

In October--November 2003 ARTEMIS-IV observed six major events on October 23 (08:24 UT), October 25 (05:20 UT), 
October 26 (06:16 UT), October 28 (11:02 UT), November 2 (a behind--the--limb event, 09:20 UT) and November 3
(09:49); Combined with data from WIND--WAVES (\cite{Bougeret}), 
these observations provide a complete view of
the radio emission induced by shock waves and electron beams from the low corona to about 1 A.U. 
Out of those, we selected three events to give an overview here.

\section{Selected Events Description}
\subsection{Event Overview}
\subsubsection{The event of October 26, 2003}
A type II with fundamental--harmonic emission, starts at 06:16 followed by high frequency type IV emission 
starting at 06:20 (high frequencies) and 06:44 (low frequencies); follows high frequency continuum emission
at 07:20 and at 08:35 UT, with considerable fine structure (cf. figure \ref{03OCT26_01}).
\subsubsection{The event of October 28, 2003}
The events start with a type II burst starting at 10:47 in ARTEMIS--IV lower frequencies, and extending into
the WIND--WAVES range. Follows a complex event, comprising an intense type III group, a type II with
fundamental--harmonic structure followed by a type IV continuum; all extend into the WIND--WAVES range.
(cf. figure \ref{03OCT28_01})
\subsubsection{The event of November 3, 2003}
The events starts at 09:49 UT with weak type III bursts, followed by  a very
strong type III group extending down to 150 kHz. Follows a type II, displaying fundamental--harmonic structure. 
There is an indication of multiple shocks since three lanes appear on the spectra; firstly a fast type II,
probably fundamental, followed by the fundamental-harmonic pair; these shocks were 
associated with a coronal wave by \cite{Vrsnak}. The event
extends into the high frequency part of the WIND--WAVES spectrum, down to about 9 MHz. A type IV 
continuum follows the type II, also extending into the WIND--WAVES band range (cf. figure \ref{03NOV03_01} upper
panels).

The SAO recordings of the type II harmonic, with high resolution, exhibit \emph{herringbone} structure, 
resulting from \emph{type III--like} emissions due to electron populations accelerated by the MHD shock.
In this recordings however the spectral shape of the \emph{herringbone} is \emph{type U--like}, thus
indicating the presence of some type of magnetic confinement for the exciter (cf. figure \ref{03NOV03_01} lower panels).
\begin{figure}[h!]
\centerline{\includegraphics[width=\textwidth,height=3cm]{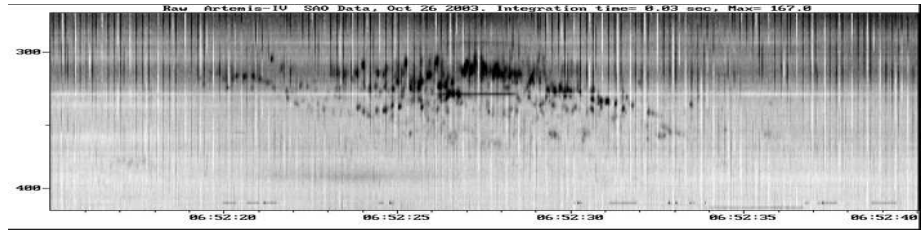}}
  \caption{ARTEMIS--IV SAO differential spectrum, in the frequency range 270--450 MHz, of Narrowband Spikes 
(26 October 2003)}
 \label{Spikes}
 \end{figure}
\begin{figure}[h!]
\centerline{\includegraphics[width=\textwidth,height=3cm]{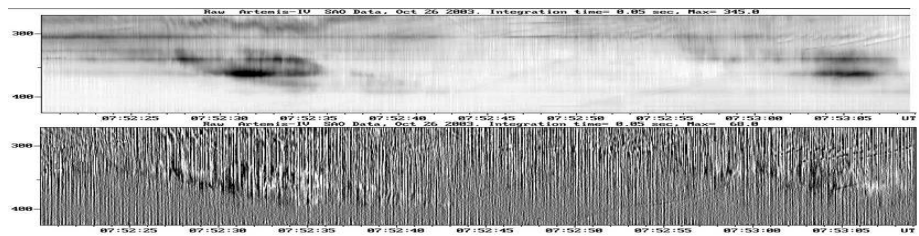}}
  \caption{ARTEMIS--IV SAO spectra, in the frequency range 350--450 MHz, of Slowly Drifting Bursts 
(cf. \cite{Jiricka} their figure 7) and fibers overlayed on a pulsating structure (26 October 2003).
Top: Intensity Spectrum, Bottom: Differential Spectrum. }
 \label{SDB}
 \end{figure}
\begin{figure}[h!]
\centerline{\includegraphics[width=\textwidth,height=3cm]{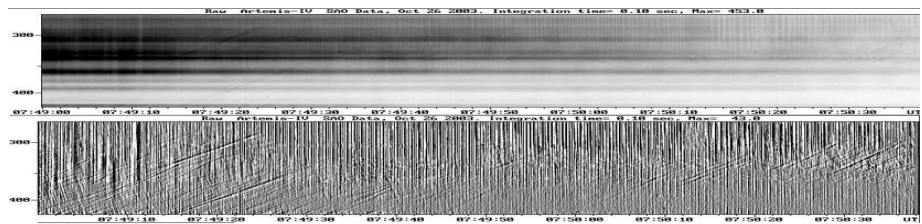}}
  \caption{ARTEMIS--IV SAO spectra, in the frequency range 270--450 MHz, of 
Pulsations and Fibers (26 October 2003). Top: Intensity Spectrum, Bottom: Differential Spectrum. }
 \label{FP01}
 \end{figure}
\begin{figure}[h!]
\centerline{\includegraphics[width=\textwidth,height=3cm]{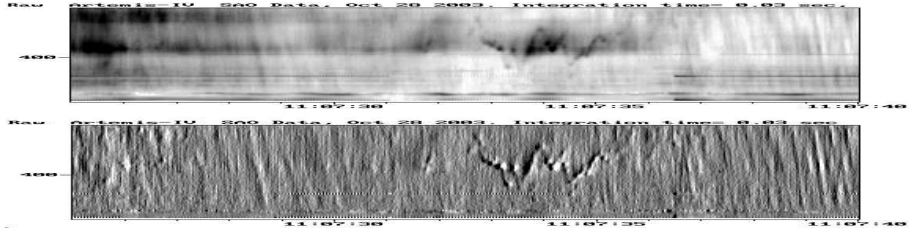}}
  \caption{ARTEMIS--IV SAO spectra (28 October 2003), in the frequency range 270--450 MHz, of a lace type structure 
(cf. \cite{Jiricka} their figure 11) overlayed with spikes. Top: Intensity Spectrum, Bottom: Differential Spectrum. }
 \label{Lace}
 \end{figure}
\begin{figure}[h!]
\centerline{\includegraphics[width=\textwidth,height=3cm]{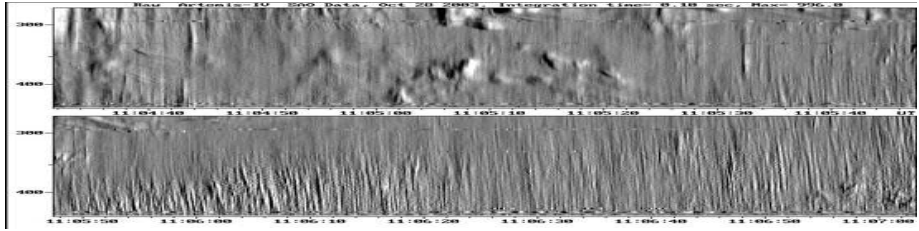}}
  \caption{ARTEMIS--IV SAO differential spectrum, in the frequency range 270--450 MHz, of Narrowband Type III 
Bursts (\cite{Jiricka} their figure 6) Pulsations and Fibers (28 October 2003). Zebra patterns are superposed
on the fibers and the pulsating structure.}
 \label{NBIII_01}
 \end{figure}
\begin{figure}[h!]
\centerline{\includegraphics[width=\textwidth,height=3cm]{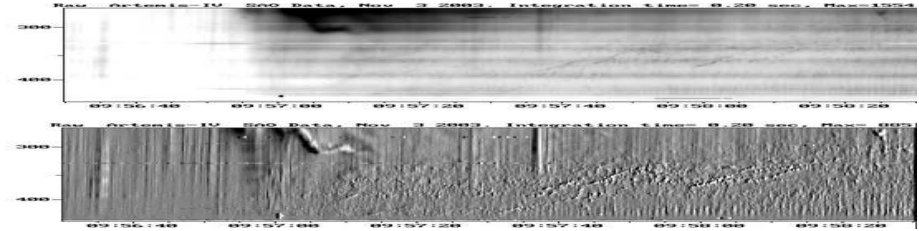}}
  \caption{ARTEMIS--IV SAO spectra (03 November 2003), in the frequency range 270--450 MHz, 
of Isolated Broadband Pulse (\cite{Jiricka} their figure 4); in the lower frequency part of the spectrum
the lower part of a pulsating structure with an EEL--type form(\cite{Chernov} their figure 1) appears. 
Top: Intensity Spectrum, Bottom: Differential Spectrum. }
 \label{FS03}
 \end{figure}

\subsection{Fine Structure}
The high sensitivity and time resolution of the SAO facilitated an examination on fine structure 
within the three studied periods; all exhibited rich fine structure embedded in the Type--IV continua.
In our analysis, the continuum background is removed by the use  of high--pass filtering on the
dynamic spectra (differential spectra in this case).

We present certain examples, which are divided according to a published morphological 
classification scheme (\cite{Jiricka}) based on Ondrejov Radiospectrograph recordings in the
0.8--2.0 GHz range. In our recordings we have detected narrow band spikes (figure \ref{Spikes}),
Slowly Drifting Bursts (figure \ref{SDB}), fibers (figures \ref{SDB},\ref{FP01}), Narrowband Type III
Bursts (figure \ref{NBIII_01}), laces (figure \ref{Lace}), zebra patterns (figure \ref{NBIII_01}),
and isolated broadband pulses (figure \ref{FS03}).
An EEL type emission (\cite{Chernov}) was found to border a pulsating structure (figure \ref{FS03}).

\section{DISCUSSION \& CONCLUSIONS}
The ARTEMIS--IV radio--spectrograph, operating in the range of 650--20 MHz, observed 6 complex events during
the super--active period of October--November 2003; three of them were presented here.
WIND-WAVES data complement nicely the ARTEMIS data and trace the radio emission from the
middle corona all the way to almost 1 A.U. The high resolution SAO recordings on the other hand
reveal a variety of fine structure which, almost, matches the comprehensive Ondrejov Catalogue 
(\cite{Jiricka}). This last, although it refers to the spectral range 0.8--2 GHz, seems to produce
similar fine structure with the metric range. 

Further analysis of the metric radio bursts fine structure is in progress.\\

\emph{This work was supported in part by the Greek Secretariat for Research and Technology through the program
Pythagoras II. The LASCO CME catalogue is generated and maintained by the Center for Solar Physics
and Space Weather, The Catholic University of America in cooperation
with the Naval Research Laboratory and NASA. SOHO is a project of international cooperation between ESA and NASA. 
The Solar Geophysical Data Catalogue is compiled and maintained by the US Department of Commerce.}

\end{document}